# ITERATIVE SOLUTION OF THE ORNSTEIN-ZERNIKE EQUATION WITH VARIOUS CLOSURES USING VECTOR EXTRAPOLATION

Herbert H. H. Homeier [1], Sebastian Rast [2], Hartmut Krienke [3]

*Institut für Physikalische und Theoretische Chemie, Universität Regensburg, D-93040 Regensburg, Germany*

The solution of the Ornstein-Zernike equation with various closure approximations is studied. This problem is rewritten as an integral equation that can be solved iteratively on a grid. The convergence of the fixed point iterations is relatively slow. We consider transformations of the sequence of solution vectors using non-linear sequence transformations, so-called *vector extrapolation processes*. An example is the vector $\mathcal{J}$ transformation. The transformed vector sequences turn out to converge considerably faster than the original sequences.

## 1 Classical Many-Particle Systems

In this paper we investigate acceleration methods for solving the fundamental equation for the pair distribution function of classical many-particle systems, the so-called Ornstein-Zernike equation. The thermodynamic properties of such systems are determined by the interaction between the particles from which the system is built up. If one knows the two-particle distribution function $g$, one can calculate all thermodynamic properties of the considered system. $g$ is defined in the canonical ensemble by [1, Chapter 4]

$$g(\mathbf{r}_1,\mathbf{r}_2) = V^2 \, \frac{\displaystyle\int \mathrm{e}^{-\beta U(\mathbf{r}_1,\ldots,\mathbf{r}_n)} d\tau(\mathbf{r}_3)\ldots d\tau(\mathbf{r}_n)}{\displaystyle\int \mathrm{e}^{-\beta U(\mathbf{r}_1,\ldots,\mathbf{r}_n)} d\tau(\mathbf{r}_1)\ldots d\tau(\mathbf{r}_n)}, \tag{1}$$

[1] na.hhomeier@na-net.ornl.gov
[2] Sebastian.Rast@chemie.uni-regensburg.de
[3] Hartmut.Krienke@chemie.uni-regensburg.de

where $V$ is the volume of the system, $n$ is the number of particles, and as usual $\beta = (k_{\rm B}T)^{-1}$, where $k_{\rm B}$ is Boltzmann's constant and $T$ is the absolute temperature. Note that the denominator is the classical configuration integral. We restrict our attention to pair potentials $u$, i.e.,

$$U(\mathbf{r}_1, \ldots, \mathbf{r}_n) = \sum_{i<j}^{n} u(\mathbf{r}_i, \mathbf{r}_j)\,. \tag{2}$$

For simplicity we consider in a first step only systems with radially symmetric interactions between identical particles. For the theoretical development (see e.g. [1, Chapters 6, 7]) of the equations it is useful to define the Mayer $f$-function:

$$f(r) := \mathrm{e}^{-\beta u(r)} - 1\,, \tag{3}$$

where $u(r)$ is the potential energy between particle 1 and 2 at distance $r$. The latter is said to be regular (short ranged) (see [2, p. 72]) if it is bounded below and satisfies

$$\int \left|\mathrm{e}^{-\beta u(r)} - 1\right| \, d\tau(\mathbf{r}) < \infty \quad \forall \quad \beta > 0. \tag{4}$$

Of this type are for example the Lennard-Jones (LJ) potential

$$u(r) = 4 \cdot \varepsilon \cdot \left( \left(\frac{\sigma}{r}\right)^{12} - \left(\frac{\sigma}{r}\right)^{6} \right), \tag{5}$$

where $\sigma$ is a distance parameter and $\varepsilon$ is the depth of the potential, and the hard sphere potential

$$u(r) = \begin{cases} \infty \;\forall\; r < \sigma \\ 0 \;\forall\; r \geq \sigma \end{cases}. \tag{6}$$

On the other hand, there are pair potentials $u(r)$ which do not obey relation (4). Nevertheless, they lead to thermodynamical behavior of systems of particles interacting with such $u(r)$. A famous example is the classical one-component plasma (OCP) with the pair potential

$$u(r) = \Gamma_p \, k_{\rm B}T \, \frac{a}{r}\,, \quad \Gamma_p = \frac{(Ze_0)^2}{4\pi\epsilon_r\epsilon_0 k_{\rm B}Ta}\,, \quad a := \left(\frac{3}{4\pi\rho}\right)^{1/3} \tag{7}$$

for particles of charge $Z$ in a neutralizing background. Here, $e_0$ is the absolute value of the elementary charge, $\epsilon_0$ is the dielectric constant, $\epsilon_r$ the relative permittivity, and $\rho$ is the average number density that can be used to define a length scale $a$. The plasma parameter $\Gamma_p$ is a dimensionless quantity. For further convenience we divide the potential in a long-range part $u^{(l)}(r)$ and a short-range part $u^{(s)}(r)$ in the following manner according to [3]:

$$u^{(l)}(r) := \Gamma_p \, k_{\mathrm{B}} T \, \frac{a}{r} \, \mathrm{erf}\,(\alpha r) \ , \quad u^{(s)}(r) := u(r) - u^{(l)}(r) \ , \tag{8}$$

where $\alpha$ is a parameter to be chosen (usually $\alpha = 1.08/a$, see [3]). For the definition of the error function $\mathrm{erf}(x)$ see [4, Chapter 7]. The Fourier transform of $u^{(l)}(r)$ can – similarly to the Fourier transform of the Coulomb potential – be calculated in the distributional sense. It is short ranged and given by

$$\tilde{u}^{(l)}(k) = 4\pi\Gamma_p \, k_{\mathrm{B}} T \, \frac{a}{k^2} \, \exp(-\frac{k^2}{4\alpha^2}) \, . \tag{9}$$

If the potential $u$ is radially symmetric and therefore only a function of $r := |\mathbf{r}_1 - \mathbf{r}_2|$ we can establish the pair distribution function $g$ as a function of $r$. To determine this quantity we are using the Ornstein-Zernike (OZ) equation [1]:

$$h = c + \rho \cdot c * h \tag{10}$$

where $*$ denotes a convolution defined by

$$[f * g](\mathbf{r}) = \int f(\mathbf{r} - \mathbf{r}')g(\mathbf{r}') \, d\tau(\mathbf{r}') \, . \tag{11}$$

The density $\rho$ is the average number density. The function $h(r) := g(r) - 1$ is called the total correlation function and $c(r)$ the direct correlation function. We note that the convolution of two radially symmetric functions is again a radially symmetric function. For the two unknown functions $h$ and $c$ we need a second equation, which is called the closure of the OZ equation and is given in general by [1]:

$$g(r) = \exp(-\beta \, u(r) + h(r) - c(r) + E(r)) \, . \tag{12}$$

$E$ is an infinite sum of multicenter integrals, the so called bridge diagrams, which are known in principle as complicated multidimensional integrals. These are very hard to evaluate. Thus, usually various simple approximations are used for them. $E(r) = 0$ is the *HyperNetted Chain approximation* or HNC closure [1], $E(r) = \ln(1 + h(r) - c(r)) - h(r) + c(r)$ is the Percus-Yevick (PY) approximation [1]. For hard spheres, Labík and Malijevský [5] introduced a semiempirical approximation (LM) of $E$. It reproduces Monte Carlo experiments excellently. There are other approximations as the very successful closure of Martynov and Sarkisov (MS) [6], where $E(r) = \sqrt{1 + 2(h(r) - c(r))} - h(r) + c(r) - 1$. For detailed formulas of the closures as we used them in our programs see below.

Together with the closure, the OZ equation is a non-linear integral equation, which can be solved in general only numerically. For hard spheres in PY approximation there is an analytic solution, too (see [7,8]).

## 2 The Direct Iteration Algorithm

The easiest algorithm for solving the OZ equation with a given closure is direct iteration using fast Fourier transformation. Due to the convolution theorem we have the following equation in $\mathbf{k}$-space:

$$\tilde{h}(k) = \tilde{c}(k) + \rho \cdot \tilde{c}(k) \cdot \tilde{h}(k), \quad k = |\mathbf{k}| \, . \tag{13}$$

The Fourier transforms are determined by the Fourier-Bessel transformation in the case of radially symmetric $f(r)$:

$$\tilde{f}(k) = 4\pi \cdot \int_0^\infty f(r) \frac{\sin{(kr)}}{kr} r^2 \, dr, \quad f(r) = \frac{1}{2\pi^2} \cdot \int_0^\infty \tilde{f}(k) \frac{\sin{(kr)}}{kr} k^2 \, dk \, . \tag{14}$$

Introducing $F(r) := f(r) \cdot r$ and $\tilde{F}(k) = k \cdot \tilde{f}(k)$ for $f = c, h$ one gets the Fourier sine transformation

$$\tilde{F}(k) = 4\pi \cdot \int_0^\infty F(r) \sin(kr) \, dr, \qquad F(r) = \frac{1}{2\pi^2} \cdot \int_0^\infty \tilde{F}(k) \sin(kr) \, dk \, . \tag{15}$$

Multiplying equation (13) by $k^2$ and introducing $\Gamma := H - C$ one obtains

$$\tilde{\Gamma} = \frac{\rho \cdot \tilde{C}^2}{k - \rho \tilde{C}} \, . \tag{16}$$

The closures can be written also in terms of $c(r) = C(r)/r$, considered as a functional $c[\gamma]$ of $\gamma(r) := \Gamma(r)/r$, and the Mayer function $f$ (see previous

section):

$$
\begin{aligned}
&\text{HNC: } C(r) = r \cdot c(r) = r \cdot (f(r)+1) \cdot \mathrm{e}^{\Gamma(r)/r} - \Gamma(r) - r\,, \\
&\text{PY:} \quad C(r) = (f(r)+1) \cdot (r + \Gamma(r)) - \Gamma(r) - r = f(r) \cdot (r + \Gamma(r))\,, \\
&\text{LM:} \quad C(r) = r \cdot (f(r)+1) \cdot \mathrm{e}^{\Gamma(r)/r + E_{\mathrm{LM}}(r)} - \Gamma(r) - r\,, \\
&\text{MS:} \quad C(r) = r \cdot (f(r)+1) \cdot \mathrm{e}^{\sqrt{1+2\Gamma(r)/r}-1} - \Gamma(r) - r\,.
\end{aligned}
\tag{17}
$$

$E_{\mathrm{LM}}(r)$ is the bridge function of Labík and Malijevský [5].

In the case of classical one-component plasmas (see previous section) we have to use a somewhat different equation from Eq. (16) because of the long-range potential involved. Following the method of Ng [3] we obtain

$$
\tilde{\Gamma}(k) = \frac{k\left(\tilde{C}^{(\mathrm{s})}(k) - k\tilde{u}^{(\mathrm{l})}(k)\right)}{k - \rho\left(\tilde{C}^{(\mathrm{s})}(k) - k\tilde{u}^{(\mathrm{l})}(k)\right)} - \tilde{C}^{(\mathrm{s})}(k)\,. \tag{18}
$$

Here, $\tilde{u}^{(\mathrm{l})}$ is the Fourier transform of the long-range part of the pair potential as defined in the previous section. $\tilde{C}^{(\mathrm{s})}$ is the Fourier transform of the short-range part of the direct correlation function multiplied by $k$. The explicit relation to $\Gamma$ is dependent on the closure. Here, we use only the HNC closure without any bridge function which is known to yield fairly good results in the region of plasma parameters $\Gamma_p$ used here [3]. Then, $C^{(s)}(r)$ is given by

$$
\text{HNC: } C^{(\mathrm{s})}(r) = r\,\exp\left(-\beta u^{(\mathrm{s})}(r) + \Gamma(r)/r\right) - \Gamma(r) - r \tag{19}
$$

where $u^{(\mathrm{s})}$ is the short-range part of the pair potential as defined in the previous section. Here, $\Gamma$ is given by $\Gamma = H - C^{(\mathrm{s})}$, so that the pair distribution function is $g(r) = (\Gamma(r) + C^{(\mathrm{s})}(r))/r + 1$.

Equation (16) together with any particular closure defined in Eq. (17) define certain integral equations. Also, equations (18) and (19) together define a further integral equation. The solution of any of these equations can be considered as a fixed point problem for the unknown function $\Gamma$. The integral equations are solved on a grid of equidistant points. Then, we put $\tilde{F}_j := \tilde{F}(j \cdot \Delta k)$, $F_j := F(j \cdot \Delta r)$, $\Delta r \cdot \Delta k = \pi/M$, where $M$ is the number of points desired for calculating the former integrals. Equation (15) for the Fourier sine transformation and its inversion becomes [9]

$$
\tilde{F}_j = 4\pi\Delta r \cdot \sum_{i=1}^{M-1} F_i \cdot \sin(ij\frac{\pi}{M})\,, \qquad F_j = \frac{\Delta k}{2\pi^2} \cdot \sum_{i=1}^{M-1} \tilde{F}_j \cdot \sin(ij\frac{\pi}{M})\,. \tag{20}
$$

Therefore, we can establish the following algorithm: Choose a $\Gamma^{(0)}$(=0 for example) for a set of equidistant $r$ and insert it in the closure getting $C^{(0)}$. This can be transformed by (20) and inserted in (16) or (18) getting a $\tilde{\Gamma}^{(1)}$, from which one gets $\Gamma^{(1)}$ by the inversion formula. This $\Gamma^{(1)}$ can be used as a new input in the iteration process. This is done until self consistency is achieved, i.e. until for a given convergence threshold $\eta > 0$ we have

$$\zeta^2 := \sum_{i=1}^{M} \left( \Gamma_i^{(j)} - \Gamma_i^{(j-1)} \right)^2 < \eta^2 \,. \tag{21}$$

The time consuming steps are the transformations of $C$ and $\tilde{\Gamma}$, so that it is desirable to reduce the number of required iterations. There are usually 200 to 1000 iterations performed until $\zeta < 10^{-10}$. Therefore the aim is to use an acceleration method for the vector sequence $\Gamma^{(j)} = (\Gamma_1^{(j)}, \Gamma_2^{(j)}, \ldots, \Gamma_M^{(j)})$ of the discretized function $\Gamma$.

## 3 Vector Extrapolation for Fixed Point Iterations

The iterative solution of systems of nonlinear equations like the OZ equation (10) with some closure of the form of Eq. (12) can often be regarded as fixed point problems

$$X = \Psi(X) \tag{22}$$

with a parameter vector $X$. In the OZ case, this vector corresponds to $\Gamma$ as discussed in the previous section. Such fixed point problems are often solved via direct iteration (Picard iteration)

$$X_0 \,, X_1 = \Psi(X_0) \,, \ldots, X_{n+1} = \Psi(X_n) \,, \ldots . \tag{23}$$

In this way, a sequence of vectors $X_n$ is generated. This sequence may or may not converge, and if it converges, it may or may not converge sufficiently fast.

Especially for slowly convergent iteration sequences $X_n$, one would like to be able to accelerate the convergence by some mathematical algorithms. Fortunately, this is possible. One may use *vector extrapolation algorithms*. These algorithms are a rapidly expanding field of mathematics. A good introduction to it is given in Chapter 4 of the textbook by Brezinski and Redivo Zaglia [10].

We discuss some general features of the acceleration of slowly convergent sequences $\{s_n\}$. Here, the sequence elements $s_n$ can be numbers, vectors, ma-trices *et cetera*. The basic principle is to use structural information hidden in the data. Once one has identified this structural information, it can be used to compute the limit faster. Usually, the result is a *sequence transformation*

$$
\begin{aligned}
& s_n \Longrightarrow t_n \\
& \{s_n\}_{n=0}^{\infty}\text{: original sequence,} \\
& \{t_n\}_{n=0}^{\infty}\text{: transformed sequence.}
\end{aligned}
\tag{24}
$$

The transformed sequence converges hopefully in a faster way.

The problems are to find a way to identify the type of structural information, and further, to construct the sequence transformation from this information. In order to discuss these problems, we introduce the notion of a *remainder* $r_n$ defined by

$$
s_n = s + r_n \,, \qquad s = \lim_{n\to\infty} s_n \,. \tag{25}
$$

Both problems are usually treated together by using a *model sequence approach*. There, one takes models for the remainder $r_n$. Then, one seeks transformations which allow – for the resulting model sequences – the exact calculation of the limit.

Thus, in this approach, one considers model sequences $\{\sigma_n\}$ of the form

$$
\sigma_n = \sigma + m_n(c_i, p_i) \quad \underset{\text{exact}}{\overset{T}{\Longrightarrow}} \quad \sigma = T_n(\sigma_n, \ldots, \sigma_{n+k}|p_i) \tag{26}
$$

Here, the model $m_n$ depends on a finite number of coefficients $c_i$, and on further parameters $p_i$. The transformation $T$ eliminates the coefficients $c_i$ and allows to calculate exactly the limit $\sigma$ of the model sequence $\{\sigma_n\}$ as function of some finite number of sequence elements $\sigma_{n+j}$. The transformation $T$ is specific for the model and depends parametric on the $p_i$.

The transformation $T$ can also be applied to the problem sequence $s_n$. Then, a sequence transformation is obtained:

$$
t_n = T_n(s_n, \ldots, s_{n+k}|p_i) \qquad \text{(approximate)} \,. \tag{27}
$$

The expectation is that the transformed sequence $\{t_n\}$ converges faster than the original sequence $\{s_n\}$ for problems that are in some sense close to the

model:

$$s_n \approx \sigma_n \; . \tag{28}$$

A useful more special model is to factor the model remainder $m_n$ into a remainder estimate $\omega_n \neq 0$ and a correction factor $\mu_n(c_i, \pi_i)$ according to

$$\sigma_n = \sigma + \omega_n \mu_n(c_i, \pi_i) \; . \tag{29}$$

The parameters then are the $\pi_i$. The remainder estimates $\omega_n$ can also be regarded as parameters. However, it is often useful to allow that the $\omega_n$ depend also on the problem sequence. This is for instance the case for Levin's remainder estimates [11] for which we display the following variants:

$$\begin{aligned} \text{“t variant”:} \quad \omega_n &= \Delta s_{n-1} = s_n - s_{n-1} \; ; \\ \text{“u variant”:} \quad \omega_n &= (n+1)\Delta s_{n-1} = (n+1)(s_n - s_{n-1}) \; . \end{aligned} \tag{30}$$

Here, and in the following, $\Delta$ denotes a difference operator acting on $n$ in the form

$$\Delta f_n = f_{n+1} - f_n \; . \tag{31}$$

Both t and u variants can be shown to be – up to some constant factor – good estimates of the true remainder $r_n$ for large classes of sequences. In this way, models allow to make use of structural information.

A further successful approach for the construction of sequence transformations is to use some simple basic sequence transformation $T_0$ iteratively:

$$s_n \xrightarrow{T_0} s_n' \xrightarrow{T_0} s_n'' \xrightarrow{T_0} \ldots \xrightarrow{T_0} s_n^{(k)} \tag{32}$$

This concept proved to be very successful in the case of scalar sequences [12–16].

In the vector case, one may take as the basic transformation $T_0$ the transformation

$$S_n' = S_{n+1} - \Omega_{n+1} \left\{ \frac{(\Delta\Omega_n, \Delta S_n)}{(\Delta\Omega_n, \Delta\Omega_n)} \right\} \; . \tag{33}$$

Here and in the following, quantities in capital letters like $S_n$, $\Omega_n$, $S_m^{(l)}$, $\Omega_m^{(l)}$ denote vectors, and $(.,.)$ denotes the usual scalar product of two vectors. The basic transformation (33) is exact for model sequences of the form

$$\Sigma_n = \Sigma + c\,\Omega_n \tag{34}$$

depending on arbitrary constants $c$. To apply (33) iteratively, one may take the same remainder estimates $\Omega_n$ in each iteration. However, it is much better [15,16] to calculate also new remainder estimates $\Omega'_n$ after each iteration step in a hierarchical consistent [15] way. Then, one obtains the *vector $\mathcal{J}$ transformation* as an important example that will be used in the sequel. It is a special case of the matrix $\mathcal{J}$ transformation that was introduced in [17]. The vector $\mathcal{J}$ transformation is defined by the recursive scheme

$$\begin{aligned}
&S_n^{(0)} = S_n; \qquad \Omega_n^{(0)} = \Omega_n; \\
&S_n^{(k+1)} = S_{n+1}^{(k)} - \Omega_{n+1}^{(k)} \left\{ \frac{\left(\Delta\Omega_n^{(k)}, \Delta S_n^{(k)}\right)}{\left(\Delta\Omega_n^{(k)}, \Delta\Omega_n^{(k)}\right)} \right\}; \\
&\Omega_n^{(k+1)} = -\Omega_{n+1}^{(k)} \left\{ \frac{\left(\Delta\Omega_n^{(k)}, \Omega_n^{(k)}\right)}{\left(\Delta\Omega_n^{(k)}, \Delta\Omega_n^{(k)}\right)} \delta_n^{(k)} \right\}; \\
&\mathcal{J}_n^{(k)}(\{S_n\}, \{\Omega_n\}, \{\delta_n^{(k)}\}) = S_n^{(k)}\,.
\end{aligned} \tag{35}$$

The $\delta_m^{(l)}$ are numbers. They correspond to the parameters $\pi_i$ in Eq. (29). Since they are numbers, the terms in curly braces in the recursive scheme (35) are also scalar numbers.

The vector $\mathcal{J}$ transformation is a straightforward generalization of the scalar $\mathcal{J}$ transformation introduced in [14] that was studied intensively in [15,16,18].

We note that the vector $\mathcal{J}$ transformation is closely related to the vector E algorithm and some projection methods (See [10]). The $\mathcal{J}$ transformations are based on iteration. On the other hand, also remainder estimates are used for the $\mathcal{J}$ transformation, and in the scalar case, it is known for which model sequences it is exact. Hence, the $\mathcal{J}$ transformations combine features of both the general schemes described above.

The problem now is how to combine vector extrapolation with Picard iteration for the solution of the fixed point equation (22). We describe a general method called *cycling* (cf. [10, p. 316]). It may be applied not only for the vector $\mathcal{J}$ transformation but for some general vector extrapolation algorithm $\mathcal{T}$. In this method, short Picard sequences are calculated from some starting value. Then extrapolation is used to calculate a new starting value for the next Picard sequence, and so on, until convergence is achieved. This is described in more detail in the following.

Thus, one constructs a vector sequence $S_n$ from some starting vector $Y$ by Picard iterations (cf. Eq. (23)) and uses extrapolation to compute a new starting vector $Y'$ in the following way:

$$\begin{aligned} S_0 &= Y \\ S_1 &= \Psi(S_0) \\ \vdots & \quad \vdots \\ S_m &= \Psi(S_{k-1}) \\ Y' &= \mathcal{T}(S_0, S_1, \ldots, S_m) \end{aligned} \tag{36}$$

This will be called an $m$-cycle. In this way, direct iteration is interspersed with extrapolation steps that provide (hopefully better) starting values for the direct iteration. Also, a number $d$ of direct iterations is normally done before cycling starts. Hence, the generated sequence of approximations is

$$\begin{array}{llll} Y_0, & Y_1 = \Psi(Y_0), \ldots & Y_d = \Psi(Y_{d-1}) & \text{direct} \\ S_0 = Y_d, & S_1 = \Psi(S_0), \ldots & Y_{d+1} = \mathcal{T}(S_0, \ldots, S_m) & \text{cycle 1} \\ S_0 = Y_{d+1}, & S_1 = \Psi(S_0), \ldots & Y_{d+2} = \mathcal{T}(S_0, \ldots, S_m) & \text{cycle 2} \\ \vdots & \vdots & \vdots & \vdots \\ S_0 = Y_{d+\nu-1}, & S_1 = \Psi(S_0), \ldots & Y_{d+\nu} = \mathcal{T}(S_0, \ldots, S_m) & \text{cycle } \nu \\ \vdots & \vdots & \vdots & \vdots \end{array} \tag{37}$$

Thus, in each cycle a new direct iteration is started form the extrapolation result of the previous cycle. After the performance of cycle no. $\nu$, the iteration function $\Psi$ has been called $N = d + \nu\, m$ times, and the extrapolation algorithm $\mathcal{T}$ has been applied $\nu$ times. Convergence checks as described in the previous section can be performed at the beginning of each cycle. As noted above, the cycling method is applied in our case by using the vector $\mathcal{J}$ transformation as transformation $\mathcal{T}$.

In practice, one sometimes observes that for a bad starting value, there is no convergence of iteration methods. This holds also for Newton-Raphson-type approaches. Actually, in studies of deterministic models for chaotic systems based on nonlinear iteration functions such a behavior is found as a generic case. Also, one should keep in mind that nonlinear iteration functions can exhibit rather peculiar behavior like fractal boundaries of the basins of attraction or strange attractors. A further remark is that the choice of the nonlinear iteration function determines whether the fixed point is stable or unstable. For instance, there are cases where Picard iterations do not converge, but Newton-Raphson methods do since then a different iteration function is used. How does this relate to vector extrapolation?

A first observation is that in principle, *nonlinear* extrapolation algorithms can also give rise to chaotic phenomena when the output is used iteratively as new input. In many applications, however, this is not the usual mode of operation and hence, it seems that such behavior has not been reported in the literature. We remark that for fixed point iterations, the combination of the direct iteration function $\Psi$ with cycling of a vector extrapolation method can be considered as a new iteration function. For instance, in the case of a 2-cycle, this function is $\Psi'(Y) = \mathcal{T}(Y, \Psi(Y), \Psi(\Psi(Y)))$. Hence, the existence of chaotic phenomena for this new function is to be expected if it is nonlinear. This is normally the case if the old iteration function or the vector extrapolation algorithm are nonlinear. However, it seems that for the new iteration function, chaotic behavior is less probable. Put another way, it is expected that an unstable fixed point of $\Psi$ can become a stable fixed point of $\Psi'$. An example is given below.

This is related to a second, and more important feature of extrapolation algorithms. They can transform divergent sequences into convergent ones. For instance, it is well-known that scalar algorithms can be used to construct analytic continuations for power series outside of their circle of convergence in the form of rational approximations that converge rapidly in many cases. Also, suitable algorithms are able to sum divergent series as in the case of the accurate calculation of ground state energies of anharmonic oscillators from their strongly divergent perturbation series [19].

The question arises whether this can be observed also for vector extrapolation algorithms. As will be shown in the next section, this really is the case. Thus, vector extrapolation processes offer the chance to achieve convergence even for cases where direct iteration does not converge.

## 4 Numerical Results

Several examples have been studied, as discussed below. We used a program *directit* for direct iterations without vector extrapolation, and a program $m2vj$ implementing direct iteration in combination with the u variant (see (30)) of the vector $\mathcal{J}$ transformation defined in (35) with

$$\delta_n^{(k)} = -\frac{1}{(n+1)(n+2)} \qquad (\text{for all } k)\,. \tag{38}$$

In the latter case, we used various numbers of direct iterations $d = n_{\text{offset}}$ as offset before cycling was started, and the length $m$ of the $m$-cycles was varied (cf. previous section). We note that use of $m2vj$ with $m = 0$, i.e., without cycling, means that direct iteration without extrapolation is used. The difference between the resulting vectors $\Gamma$ with and without extrapolation is given by the quantity

$$\delta = \sqrt{\sum_{i=1}^{M} \left(\Gamma_i^{directit} - \Gamma_i^{m2vj}\right)^2} \tag{39}$$

where $M$ denotes the number of points in the grid.

As a first example we consider a system of hard spheres at various densities using the PY and HNC approximation. The calculations were performed on a Sun Sparc workstation using 512 or 128 points for the function $\Gamma$ at $\Delta r = 0.01\sigma$ where $\sigma$ is the diameter of the hard sphere, respectively. The number of iterations that are needed to reach the convergence threshold $\eta$ (cf. Eq. (21)) starting from $\Gamma^{(0)} = 0$, is denoted by $N$. The results in Table 1 show, that it is possible to reduce the total number of iterations $N$ approximately by a factor of two. The acceleration effect does not depend significantly on the choice of the approximation for the bridge function $E$. There was also a variety of values of $m$ and $n_{\text{offset}}$ that yielded rather good results. As shown in Table 1, there is no significant difference between the resulting vectors $\Gamma$ obtained by direct iteration with or without acceleration, since $\delta \approx 10^{-10}$.

The same closures were also applied to Lennard-Jones systems. The results are presented in Table 2. Again, substantial reductions of the total number of iterations are possible.

We also studied the Labík and Malijevský (LM) approximation and the Martynov-Sarkisov (MS) closure for hard sphere systems and for Lennard-Jones particles. The results are presented in Table 3. For this example, we performed the calculations on an Iris Indigo of Silicon Graphics using 1024 points for the function $\Gamma$, $\Delta r = 0.005\sigma$, and the starting vector was $\Gamma^{(0)} = 0$. As noted above, the case $m = 0$ corresponds to performing only direct iterations without any acceleration.

As a further example, we present results of calculations on classical one-component plasmas using Ng renormalization [3] with HNC closure (without

Table 1
**Hard spheres**
Diameter: $\sigma = 1$, number density: $\rho^* := \rho/\sigma^{-3}$. Starting vector: $\Gamma^{(0)} = 0$, total number of iterations: $N$, grid size: $M$, threshold: $\eta = 1 \cdot 10^{-10}$ (cf. Eq. (21)), deviation: $\delta$ (cf. Eq. (39)), cycle length: $m$, offset: $n_{\text{offset}}$, calculation on Sun workstation.

| case | program | $m$ | $n_{\text{offset}}$ | $\delta$ | $N$ |
|---|---|---|---|---|---|
| A | *directit* | – | – | – | 592 |
| | *m2vj* | 8 | 14 | no convergence up to 900 iterations | |
| | *m2vj* | 9 | 14 | $7.092 \cdot 10^{-10}$ | 295 |
| | *m2vj* | 9 | 15 | $4.955 \cdot 10^{-10}$ | 306 |
| | *m2vj* | 10 | 14 | $4.935 \cdot 10^{-9}$ | 400 |
| | *m2vj* | 10 | 15 | $8.240 \cdot 10^{-10}$ | 346 |
| B | *directit* | – | – | – | 124 |
| | *m2vj* | 7 | 10 | $4.904 \cdot 10^{-10}$ | 51 |
| | *m2vj* | 8 | 5 | $5.640 \cdot 10^{-11}$ | 41 |
| | *m2vj* | 8 | 10 | $4.681 \cdot 10^{-10}$ | 47 |
| C | *directit* | – | – | – | 488 |
| | *m2vj* | 8 | 10 | $2.318 \cdot 10^{-9}$ | 209 |
| D | *directit* | – | – | – | 384 |
| | *m2vj* | 9 | 14 | $3.080 \cdot 10^{-9}$ | 455 |
| | *m2vj* | 10 | 13 | $2.716 \cdot 10^{-9}$ | 179 |
| | *m2vj* | 10 | 14 | $2.963 \cdot 10^{-9}$ | 257 |
| | *m2vj* | 11 | 13 | $3.022 \cdot 10^{-9}$ | 182 |

A: PY$^\dagger$ approximation, $\rho^* = 0.7$ ($\Delta r = 0.01\sigma$, $M = 512$)
B: PY$^\dagger$ approximation, $\rho^* = 0.4$ ($\Delta r = 0.01\sigma$, $M = 512$)
C: PY$^\dagger$ approximation, $\rho^* = 0.7$ ($\Delta r = 0.04\sigma$, $M = 128$)
D: HNC$^\dagger$ approximation, $\rho^* = 0.7$ ($\Delta r = 0.01\sigma$, $M = 512$)
$^\dagger$ The abbreviations for the closures are explained in Section 1 and Eq. (17).

taking any further bridge diagrams into account). The results are displayed in Table 4. Again, we used $M = 1024$ points in the grid and started from $\Gamma^{(0)} = 0$. It is worth noting that for higher plasma parameters $\Gamma_p$ it is advantageous to choose a higher number $n_{\text{offset}}$ of iterations without any acceleration than at low $\Gamma_p$, but the cycle length is not to be changed very much. Actually we found our best results nearly at all values of $\Gamma_p$ at $m = 10$. This is favorable because the computational costs for the extrapolation rise with increasing $m$.

Table 2
**Lennard-Jones potential**
LJ‡ parameters: $\sigma = 1$, $\beta\varepsilon = 0.5$, number density: $\rho^* = \rho/\sigma^{-3}$. Grid: $M = 512$, $\Delta r = 0.01\sigma$, threshold: $\eta = 1 \cdot 10^{-10}$, starting vector: $\Gamma^{(0)} = 0$, calculation on Sun workstation, other symbols see Table 1.

| case | program | $m$ | $n_{\text{offset}}$ | $\delta$ | $N$ |
|---|---|---|---|---|---|
| A | *directit* | – | – | – | 137 |
| | *m2vj* | 7 | 10 | $6.532 \cdot 10^{-10}$ | 51 |
| | *m2vj* | 8 | 10 | $4.950 \cdot 10^{-10}$ | 56 |
| B | *directit* | – | – | – | 958 |
| | *m2vj* | 7 | 20 | $8.174 \cdot 10^{-9}$ | 349 |
| | *m2vj* | 7 | 30 | $4.296 \cdot 10^{-9}$ | 383 |
| | *m2vj* | 8 | 20 | $4.705 \cdot 10^{-9}$ | 372 |
| | *m2vj* | 10 | 50 | $7.854 \cdot 10^{-9}$ | 480 |
| C | *directit* | – | – | – | 615 |
| | *m2vj* | 8 | 30 | $1.234 \cdot 10^{-10}$ | 517 |

A: PY† approximation, $\rho^* = 0.5$
B: PY† approximation, $\rho^* = 0.9$
C: HNC† approximation, $\rho^* = 0.9$
† The abbreviations for the closures are explained in Section 1 and Eq. (17).
‡ The Lennard-Jones potential is defined in Eq. (5).

As one can see from the tables, the acceleration is in any case successful if one performs a sufficient number of iterations (usually approximately 10) without acceleration before cycling. At higher number densities as in Table 3 it can be necessary to perform more direct iterations. Here, the best result was obtained after 100 or more direct iterations. But see in contrast the last example of Table 3 at a very high number density, where the best result is obtained using no iterations without acceleration. The length of cycles $m$ is not needed to be very high, usually 8 to 20. This is desirable because in this way storage and computing times for the cycling are reduced. The use of longer cycles can even be disadvantageous as one can see from Tables 2 or 3. For Lennard-Jones systems the results are of similar quality as in the hard sphere case (see Tables 2 and 3). Especially, note the relatively high number densities up to $\rho^* = 1.2$ in Table 3, which was considered up to now to be intractable with direct iteration [20], when starting from $\Gamma^{(0)} = 0$.

Table 3
**Hard spheres and Lennard-Jones potential**
Sphere diameter/LJ‡ parameter: $\sigma = 1$. Starting vector: $\Gamma^{(0)} = 0$, threshold: $\eta = 1 \cdot 10^{-10}$ (cf. Eq. (21)), grid size: $M = 1024$. Program: *m2vj*, calculation on Iris Indigo, other symbols see Table 1.

| case | $m$ | $n_{\text{offset}}$ | $N$ |
|---|---|---|---|
| A | 0 | – | 576 |
| | 14 | 150 | 361 |
| | 15 | 100 | 309 |
| | 16 | 120 | 308 |
| | 16 | 150 | 389 |
| B | 0 | – | 913 |
| | 9 | 0 | 611 |
| | 9 | 1 | 572 |
| | 15 | 100 | 581 |
| | 15 | 200 | 537 |
| C | 0 | – | 840 |
| | 15 | 20 | 661 |
| | 15 | 100 | 469 |
| | 15 | 120 | 681 |
| D | 0 | – | 956 |
| | 7 | 0 | 241 |
| | 7 | 1 | 378 |

A: LM† bridge function, hard spheres, $\rho^* = 0.75$ ($\Delta r = 0.005\sigma$)
B: MS† approximation, hard spheres, $\rho^* = 0.80$ ($\Delta r = 0.005\sigma$)
C: MS† approximation, Lennard-Jones‡, $\beta\varepsilon = 0.5$, $\rho^* = 0.90$ ($\Delta r = 0.005\sigma$)
D: MS† approximation, Lennard-Jones‡, $\beta\varepsilon = 0.1$, $\rho^* = 1.20$ ($\Delta r = 0.01\sigma$)
† The abbreviations for the closures are explained in Section 1 and Eq. (17).
‡ The Lennard-Jones potential is defined in Eq. (5).

In order to assess the additional costs for the extrapolation steps in the cycling algorithm, we measured the total CPU time to run our programs for several examples. As it is well-known, such measurements have to be interpreted cautiously since the results depend not only on the basic algorithm, but also on the skills of the programmer and the machine architecture and utilization when the programs are run.

Table 4
**One-component Plasmas**
Plasma parameter: $\Gamma_p$, $a = 1$ (cf. Eq. (7)). HNC† approximation. Ng renormalization, $\alpha$ (parameter in error function): $1.08/a$ (cf. Eq. (8) and [3]). Number density: $\rho = 3/(4\pi\, a^3)$. Grid: $M = 1024$, $\Delta r = 0.01\, a$. Starting vector: $\Gamma^{(0)} = 0$. Program: *m2vj*, calculation on Iris Indigo. Other symbols see Table 1.

| $\Gamma_p$ | $m$ | $n_{\text{offset}}$ | $N$ |
|---|---|---|---|
| 10 | 0 | – | 70 |
| 10 | 5 | 0 | 55 |
| 10 | 7 | 5 | 30 |
| 10 | 8 | 5 | 33 |
| 10 | 10 | 0 | 23 |
| 50 | 0 | – | 252 |
| 50 | 7 | 10 | 75 |
| 50 | 10 | 0 | 78 |
| 50 | 10 | 10 | 77 |
| 100 | 0 | – | 458 |
| 100 | 8 | 40 | 275 |
| 100 | 8 | 60 | 412 |
| 100 | 10 | 10 | 275 |
| 100 | 10 | 50 | 161 |
| 100 | 14 | 40 | 341 |
| 120 | 0 | – | 537 |
| 120 | 8 | 100 | 614 |
| 120 | 10 | 50 | 414 |
| 120 | 10 | 100 | 189 |
| 120 | 12 | 100 | 361 |

† The abbreviations for the closures are explained in Section 1 and Eq. (17).

The basic result is that for typical cycle lengths and grid sizes the costs per iteration for the extrapolation part are of the same order as the costs for the direct iteration alone. Incidentally, this shows that for the direct iteration the costs are rather low. This is due to the use of the fast Fourier transform. For more complicated iteration functions, the relative costs of the extrapolation are expected to be better. But even for the very fast direct iterations of the current application, there are cases where it is possible to reduce the total execution time by a factor of up to two by using vector extrapolation. This follows from the following examples.

The case of classical one-component plasmas is treated in Table 5. For given values of the plasma parameter $\Gamma_p$, the total time taken by direct iteration only, i.e., with $m = 0$, is compared to the total time for the case with the lowest number of total iterations in Table 4) for $m \neq 0$, i.e., including acceleration. For example, for $\Gamma_p = 120$, the CPU time for $m = 0$ is compared to the CPU time for $m = 10$ and $n_{\text{offset}} = 100$.

Table 5
**CPU Times**
Plasma parameter (cf. Eq. (7)): $\Gamma_p$. Calculations on Sun workstation

| $\Gamma_p$ | Direct iteration alone | Direct iteration + acceleration | Saving |
|---|---|---|---|
| 100 | 37.7 s | 28.0 s | 25.7 % |
| 120 | 44.1 s | 22.0 s | 50.1 % |

A similar result was obtained for the last example of Table 3: 88.5 seconds of the direct iteration versus 44.6 seconds with acceleration, a saving of 49.6 percent of the CPU time.

We want to stress that – for the proposed method – the computing times are very low. On workstations, each run can be completed in times of the order of a minute. This should be compared to computer simulations where computing times of the order of hours or more are required.

Finally, we want to give an example that shows that there are cases where vector extrapolation can be used successfully to find fixed points even when the direct iteration does not converge when started from $\Gamma^{(0)} = 0$. Actually, for this starting value, even the Newton-Raphson-type algorithm of Labík, Malijevský, and Voňka [20] does not converge. For the latter, we use a Fortran program called *lensub*.

The example is a hard-sphere system for $\rho^* = 0.85$ with LM closure. Using a grid of size $M = 512$ and with $\Delta r = 0.01\sigma$, the results are the following:

- The Newton-Raphson-type algorithm [20] does not converge for a starting vector $\Gamma^{(0)} = 0$. It does converge for better starting vectors (for instance from a run with lower density).
- The direct iteration does not converge for a starting vector $\Gamma^{(0)} = 0$. It also does not converge when the solution obtained with *lensub* is used as starting vector. This reveals that this starting vector corresponds to an unstable fixed point for the direct iteration. See also Figure 1 and 2.

– The direct iteration in combination with the vector extrapolation converges for a starting vector $\Gamma^{(0)} = 0$ to the same solution as the Newton-Raphson-type algorithm. Some examples are displayed in Table 6.

Table 6
**Hard Spheres**
Density $\rho^* = 0.85$. LM† closure. Program: *m2vj*, calculation on Iris Indigo. Grid: $M = 512$, $\Delta r = 0.01\sigma$. Threshold: $\eta = 10^{-10}$. Other symbols see Table 1.

| Example | $m$ | $n_{\text{offset}}$ | $N$ |
|---|---|---|---|
| 1 | 9 | 40 | 991 |
| 2 | 9 | 100 | 891 |
| 3 | 9 | 150 | 881 |
| 4 | 9 | 200 | 981 |

† The abbreviations for the closures are explained in Section 1 and Eq. (17).

In Figure 1, we plot semi-logarithmically the values of $\zeta$ defined in Eq. (21) for this example for the direct iteration. Thus, $\zeta$ measures the distances between consecutive vectors in the iteration.

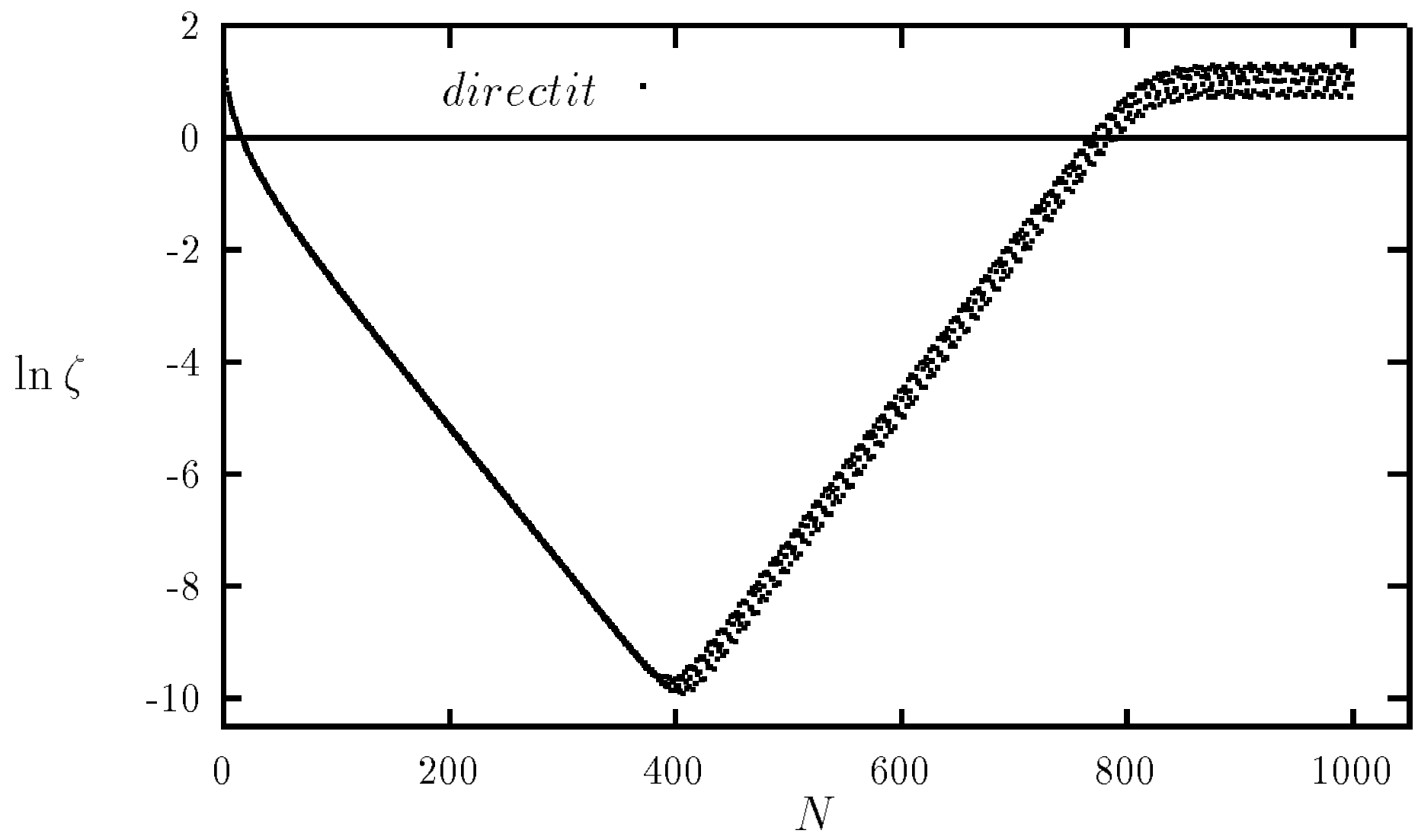

Fig. 1. Unstable Fixed Point of the Direct Iteration

It is clearly seen that the direct iteration first seems to converge but then it is starting to change rapidly until a quasiperiodic behavior is reached. The latter is displayed in an expanded representation in Figure 2.

In Figure 3, we plot the values of $\zeta$ defined in Eq. (21) for the case of Example 3 in Table 6, i.e., for the direct iteration in combination with vector extrapolation. Convergence is rather smooth apart from some step-like features.

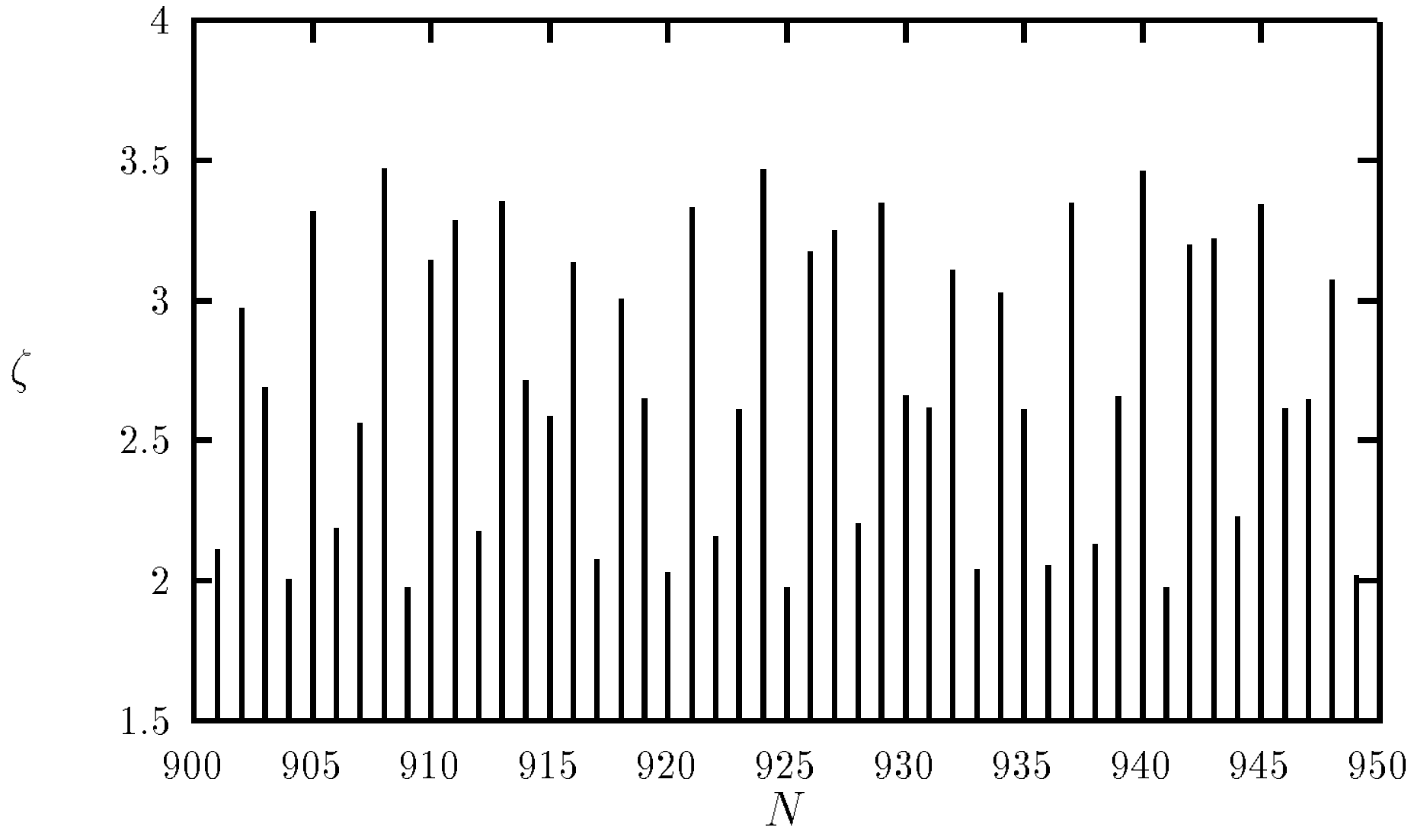

Fig. 2. Quasiperiodic Behavior of the Direct Iteration

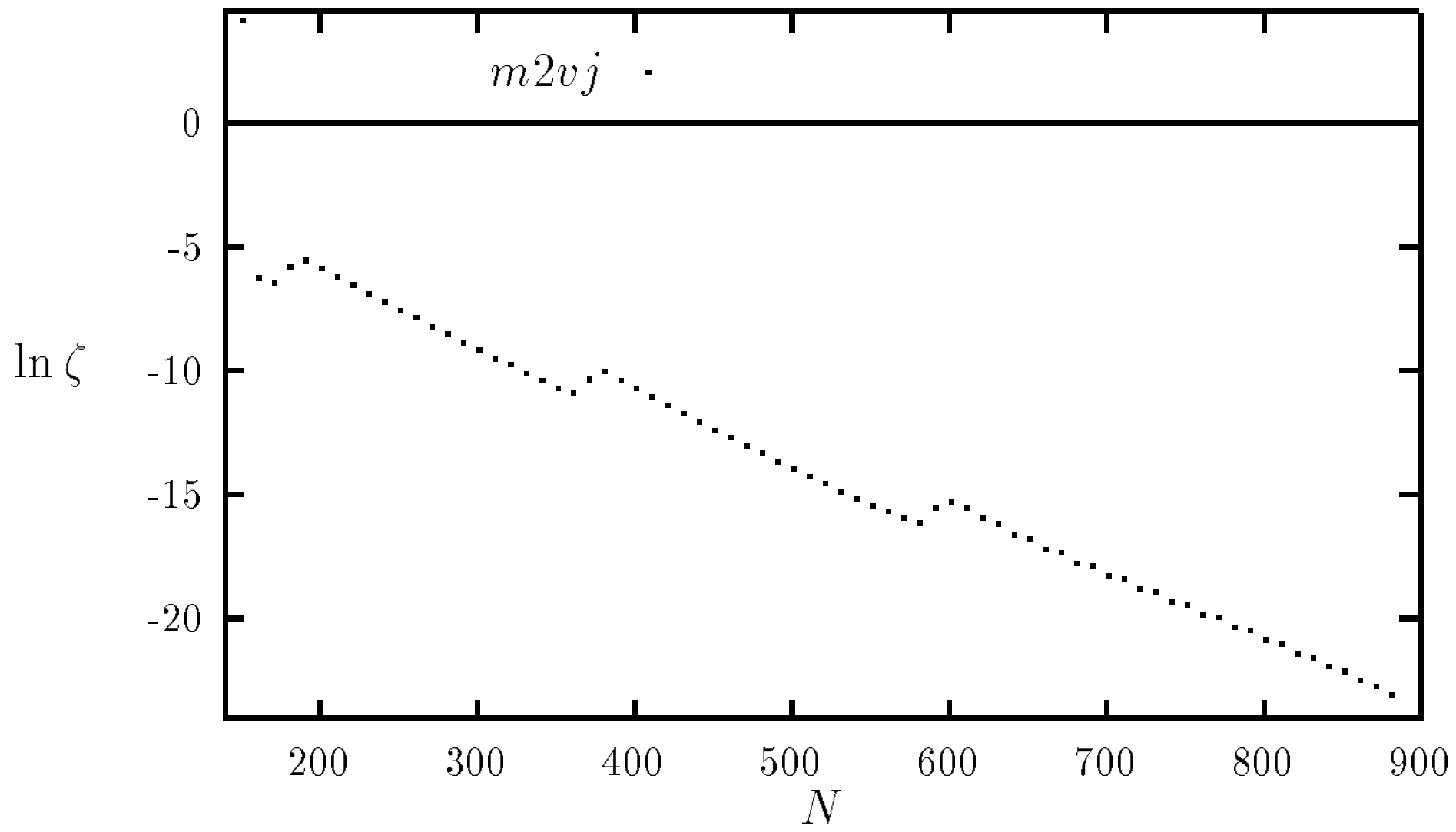

Fig. 3. Convergence of Vector-extrapolated Iteration

Before leaving this topic we note that for the same system and grid for a somewhat smaller density of $\rho^* = 0.8$ the direct iteration converges from a starting vector $\Gamma^{(0)} = 0$ while the Newton-Raphson-type algorithm of [20] does not converge using this starting vector.

An important advantage of the present algorithm is the very simple form of the direct iteration. This makes the treatment of more complicated cases as particles with dipoles or more realistic model potentials defined on the grid very easy. Essentially, one only has to reprogram the subroutine for the computation of the Mayer function. In the case of a renormalization of the OZ scheme as for OCPs, the changes are also easy to implement. All other algorithms as that due to Labík, Malijevský, and Voňka [20, p. 710] need more sophisticated programming, than the method presented here, because no calculation of the gradient of the iteration function is required as in all algorithms based on the Newton-Raphson method.

The promising results of this work suggest to study the combination of direct iteration in combination with acceleration methods also for more complicated model potentials and multicomponent systems. The applicability of further known vector extrapolation processes in the field of the Ornstein-Zernike equation should be studied. In the opinion of the authors, also the development of new powerful vector extrapolation processes is possible and desired that are tailored to speed up vector iteration processes even further. Thus, we stress that the methods of the present work still have the potential to further improvements. Finally, it is an interesting question whether it is possible to use acceleration methods profitably also in combination with more complicated algorithms like Newton-Raphson iterations. In summary, in the context of the Ornstein-Zernike equation, the combination of direct iteration methods with vector extrapolation has been shown to be a fruitful alternative to other methods.

## Acknowledgement

One of the authors (HHHH) is grateful to Prof. Dr. E. O. Steinborn for his support and the excellent working conditions at Regensburg. The authors are pleased to acknowledge the fruitful discussions with Prof. Dr. J. Barthel and his valuable comments concerning the subject of this research. Help of the staff of the computing centre of the University of Regensburg, notably of the local TEX expert M. Middleton, is thankfully acknowledged.